\documentclass[prl,aps,showpacs,twocolumn]{revtex4}

\usepackage{epsfig}
\usepackage{epsf}
\usepackage{bm}                 
\usepackage{amsmath}
\usepackage{amssymb}
\usepackage{dcolumn}
\usepackage{graphicx}

\input amssym.def
\input amssym

\newcommand{\M}{\mathfrak{M}}
\newcommand{\fL}{\mathfrak{L}}
\newcommand{\fS}{\mathfrak{S}}
\newcommand{\NL}{\mathfrak{NL}}
\begin{document}

\title{Causality, Entanglement, and
Quantum Evolution Beyond Cauchy Horizons}

\author{Ulvi Yurtsever} \email{Ulvi.Yurtsever@jpl.nasa.gov}

\affiliation{Quantum Computing Technologies Group, Jet Propulsion 
Laboratory, California Institute of Technology \\ Mail Stop 126-347, 
4800 Oak Grove Drive, Pasadena, California 91109-8099}

\author{George Hockney} \email{George.Hockney@jpl.nasa.gov}

\affiliation{Quantum Computing Technologies Group, Jet Propulsion 
Laboratory, California Institute of Technology \\ Mail Stop 126-347, 
4800 Oak Grove Drive, Pasadena, California 91109-8099}

\date{\today}

\begin{abstract}
We consider a bipartite entangled system half of which falls through the
event horizon of an evaporating black hole, while the other half remains
coherently accessible to experiments in the exterior region. Beyond
complete evaporation, the evolution of the quantum state
past the Cauchy horizon cannot remain unitary, raising the questions:
How can this evolution be described as a quantum map, and how is
causality preserved? The answers are subtle, and are linked in
unexpected ways to the fundamental laws of quantum mechanics. We show
that terrestrial experiments can be designed to constrain
exactly how these laws might be altered by evaporation.
\end{abstract}

\pacs{03.67.-a, 03.65.Ud, 04.70.Dy, 04.62.+v}

\maketitle

Standard proofs
that non-local Bell correlations~\cite{bellcorr} between
parts of an entangled system cannot be used to acausally
signal (transfer information) rely on quantum evolution being
everywhere unitary. However, as Hawking~\cite{hawkingnonunit} first
pointed out when he gave examples of non-unitary but causal
maps for evaporating black holes,
unitarity, a sufficient but not a necessary condition
for causality, may break down in the late stages
of black-hole evaporation. In this letter we ask: When entangled systems
partly cross the event horizons of evaporating black holes
(or Cauchy horizons of other, more general
naked singularities) and partly remain
coherently accessible to experiments outside, what constraints
on their non-unitary, and possibly nonlinear quantum evolution
would ensure causality? and: Can signaling (acausal) evolution be detected
at large distances if it indeed
does take place under the extreme conditions near
naked singularities and evaporating black-hole interiors?

Why expect the experimentally well-established law of unitary
evolution to break down during black-hole evaporation?
Consider, for definiteness, a pure quantum-field state
which gravitationally collapses to form an evaporating
Schwarzschild black hole (Fig.\,1).
\begin{figure}[htp]
\hspace{1.6in}
\centerline{
\input epsf
\setlength{\epsfxsize}{3.740in}
\setlength{\epsfysize}{3.220in}
\epsffile{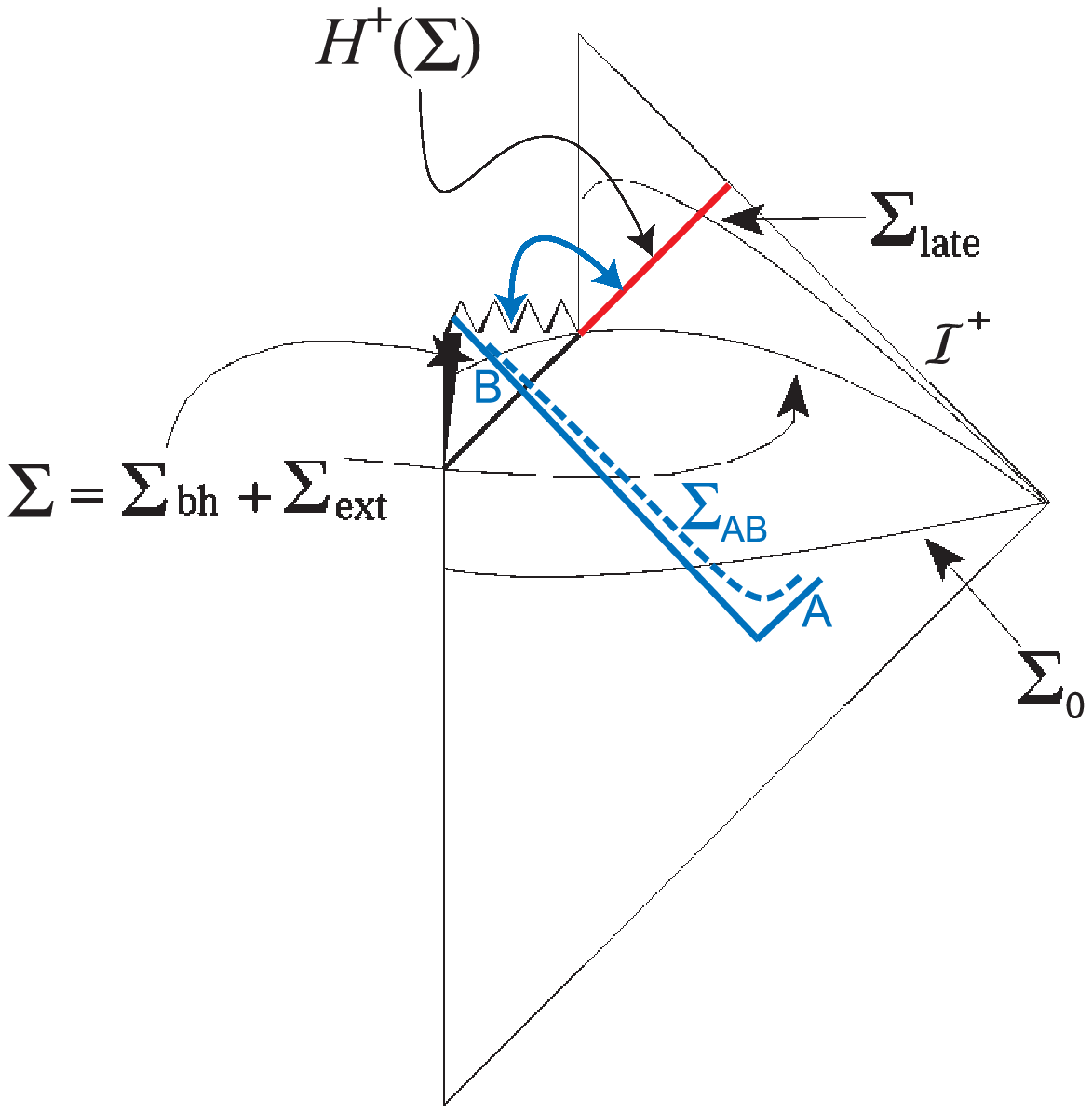}
}
\vspace{-0.7in}
\caption[figure]{\label{fig:figure1} Conformal diagram illustrating
the causal structure of a spacetime with an evaporating black hole
(vertical lines on the left depict the axes of rotational symmetry).
[The causal geometry illustrated by the superimposed blue drawing
refers to the second paragraph following Eq.\,(15) below.]
The spacelike hypersurface $\Sigma_0$
passes through the collapsing star before the black
hole has formed, $\Sigma$ is a surface through the black hole
just before it evaporates, and $\Sigma_{\rm late}$
is a surface at late times, after
complete evaporation. The red line illustrates the Cauchy horizon
$H^{+}(\Sigma )$ for $\Sigma$ or $\Sigma_0$;
it is the future null cone of the ``point" (really a singularity)
of complete evaporation.
Because evaporation is largely thermal, quantum evolution
through $H^{+}(\Sigma )$ from the time slice
$\Sigma_0$ to the slice $\Sigma_{\rm late}$
cannot be described as a unitary map.}
\vspace{-0.30in}
\end{figure}
Initially given by $| \Psi_0 \rangle$ on the
(partial) Cauchy surface $\Sigma_0$
in Fig.\,1, the state evolves unitarily
(at least in semiclassical gravity)
during and after gravitational collapse:
at any intermediate time slice $\Sigma$, it can be written as $
|\Psi_{\Sigma}\rangle = U_{\Sigma \, \Sigma_0 }
|\Psi_0 \rangle$, where $U_{\Sigma \, \Sigma_0 }$
is the unitary time evolution operator acting on the Fock space
of field states. An external observer
in the asymptotically flat region outside the event horizon has no
causal communication with the interior $\Sigma_{\rm bh}$; she
would describe the state of the quantum field by the reduced density matrix
\begin{equation}
\rho_{\rm ext} = {\rm Tr}_{\Sigma_{\rm bh}} \, |\Psi_\Sigma \rangle
\, \langle \Psi_\Sigma | \; 
\end{equation}
obtained by tracing over the interior field degrees
of freedom inside the horizon.
As the black hole settles down to a stationary
state on the time slice $\Sigma$,
the mixed state $\rho_{\rm ext}$ can be shown (via non-trivial
calculation~\cite{bhentropy}) to approach precisely a
thermal state $\rho_H$ at the Hawking temperature $T_H = \hbar c^3
/(8 \pi k_B G M)$, where $M$ is the hole's mass.
As long as the back action of
the Hawking radiation on spacetime is negligible (an eternal black hole),
matter remains in the pure state $|\Psi_{\Sigma} \rangle$,
which unitarily evolves to become entangled with its collapsed half
inside the emerging event horizon. But what happens at late times,
after this back action eventually destroys the black hole completely?
In semiclassical gravity, it is impossible
to escape the conclusion that the state $\rho_{\rm late}$ of the field
on the late time slice $\Sigma_{\rm late}$ (Fig.\,1) is mixed:
$\rho_{\rm late} \approx \rho_H$. The resulting
evolution $|\Psi_0 \rangle
\longmapsto \rho_{\rm late}$ cannot be unitary, as it maps pure states
into mixed states.
This inevitable breakdown of unitarity
can only be avoided by postulating a remnant that
persists at late times, continuing to carry the correlations ``lost"
in the state $\rho_{\rm late}$ by remaining entangled with the outgoing
Hawking radiation.

The lesson we draw is: compared to the conditions encountered in local
laboratory physics, conditions in the interiors
of evaporating black holes are so extreme that
the ordinary laws of quantum evolution
may be profoundly altered~\cite{ftnote1}.
What kinds of non-unitary quantum dynamics might govern entangled
multi-partite systems as their subsystems
cross the Cauchy horizons of evaporating black holes?
We argue that this dynamics must be probability-preserving,
it can be (generally) nonlinear, and it must be local.
The class of non-unitary maps
(``superscattering operators")
discussed by Hawking~\cite{hawkingnonunit}
is obtained via the additional constraint of linearity.
We will show that linearity (along with probability conservation
and locality) is sufficient to preserve causality~\cite{ftnote2};
acausal signaling is possible only with nonlinear maps.
Nonlinear generalizations of quantum mechanics and their
implications for measurement theory and causality have been discussed
by many authors~\cite{nlrefs}; it is not our goal in this letter to
contribute to these developments. We adopt the conservative position
that at most a minimal generalization of quantum theory---namely one that
allows for the possibility of
nonlinear quantum maps while keeping the rest of the formalism
intact---is necessary
to understand the non-standard quantum dynamics of black-hole evaporation.
There is, of course, no experimental
evidence for quantum nonlinearity
under local laboratory conditions~\cite{weinberg};
however, whether linearity
continues to hold under the extreme conditions
of evaporating black-hole interiors is a question
yet to be decided by experiment. Remarkably, a simple terrestrial
experiment can be designed to probe this question as we now discuss.

\begin{figure}[htp]
\hspace{3.0in}
\centerline{
\input epsf
\setlength{\epsfxsize}{3.850in}
\setlength{\epsfysize}{2.985in}
\epsffile{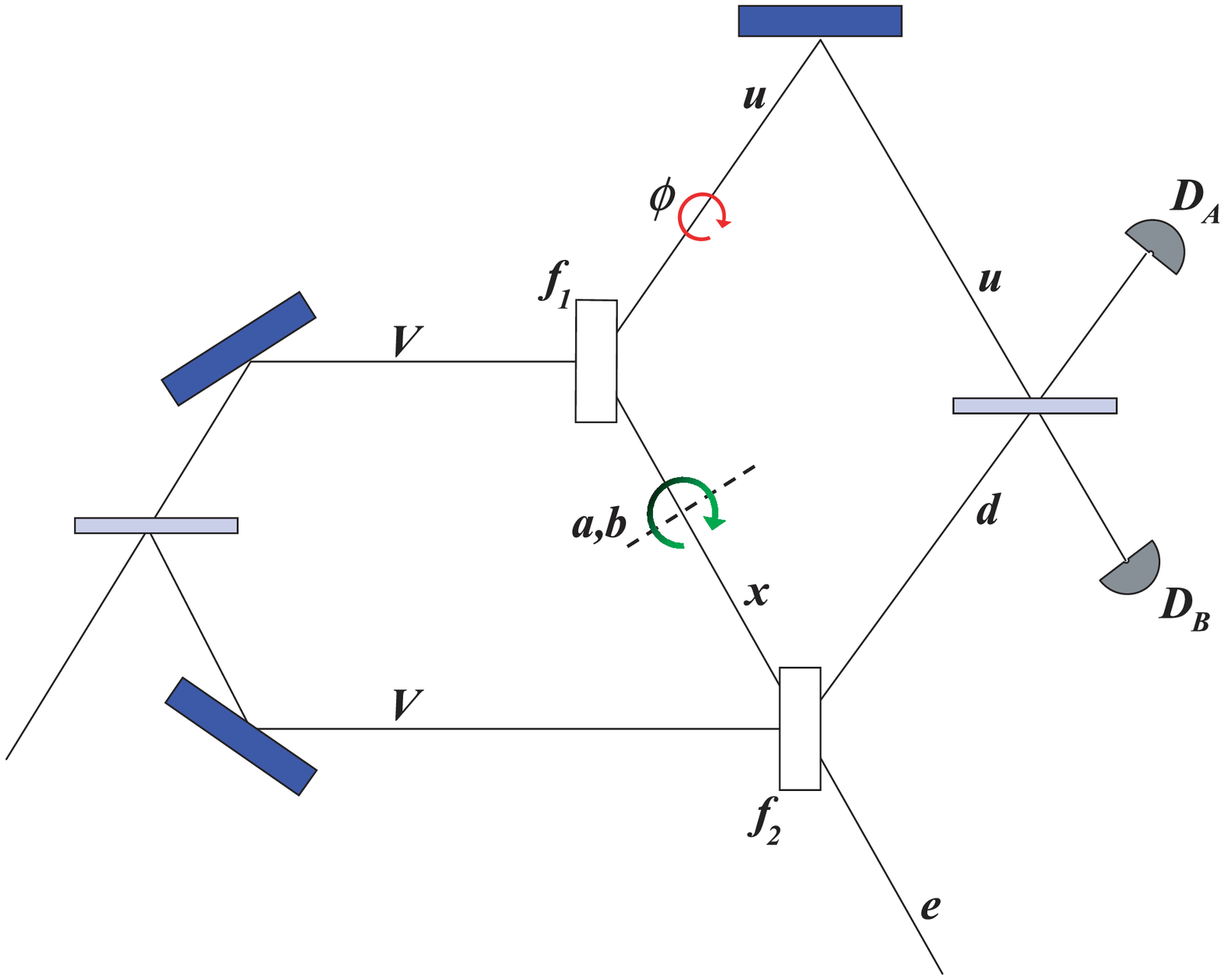}
}
\vspace{-0.40in}
\caption[figure]{\label{fig:figure2}
The Zou-Wang-Mandel interferometer~\cite{mandeletal} for the trans-horizon
Bell-correlation experiment. Gray rectangles are
50:50 beam splitters, white rectangles are the two nonlinear parametric
down-converting crystals with efficiencies $f_1$, $f_2$; blue
rectangles are mirrors. A phase delay is placed on the idler
beam labeled $u$, and an adjustable-angle
polarization rotator is placed on the signal beam $x$ which is aligned
with the second signal beam labeled $e$. Since both pump beams
are blocked past the nonlinear crystals, the output state lies
in the Hilbert
space ${\cal H}_{u} \otimes {\cal H}_{d} \otimes {\cal H}_{e}$ of
the signal and idler beams;
it is monitored by the single-photon detectors $D_A$ and $D_B$.}
\vspace{-0.20in}
\end{figure}
Consider the optical setup schematically illustrated in Fig.\,2,
a straightforward modification of a well-known Bell-correlation
experiment by Mandel et.\ al.~\cite{mandeletal} The pump beam
(typically from the output of a uv-argon laser) is split
into two beams which
interact with two separate nonlinear crystals to produce
correlated photons in two pairs of idler and signal beams, labeled
$u$, $x$, and $d$, $e$, respectively. The key feature in the
design of the experiment is the alignment of the first signal beam
$x$ with the second signal beam $e$, which makes photon number-states
in the beams (modes) $x$ and $e$ indistinguishable (in practice,
the alignment needs to be accurate only to within the transverse
laser coherence length). In the actual experiment
the first signal beam $x$ may pass through the second
nonlinear crystal as a consequence of its alignment with $e$,
but its probability of further down-conversion,
proportional to $|V f_1 f_2 |^2$, is negligible since $|f_i |\ll 1$
and $|V f_i |\ll 1$, $i=1, \; 2$, where $V$ is the dimensionless
amplitude of each of the two pump-beam pulses
(photon number $\propto |V|^2$).
We shall assume that both nonlinear crystals produce down-converted
photons in a fixed (linear) polarization state.
The quantum state output by this
configuration belongs to the Hilbert space ${\cal H} \equiv {\cal H}_u
\otimes {\cal H}_d \otimes {\cal H}_e$, where the ``up" and
``down" idler-beam Hilbert spaces are generated by the
orthonormal basis states
\begin{equation}
{\cal H}_u \equiv < \{ | 0 \rangle_u , \; |1 \rangle_u \} > \; ,
\;\;\;\; {\cal H}_d \equiv < \{ | 0 \rangle_d , \; |1 \rangle_d \} >
\; ,
\end{equation}
and the ``escaping" signal-beam Hilbert space is generated by the
basis states
\begin{equation}
{\cal H}_e \equiv < \{ | 0 \rangle_e , \; |1 \rangle_e
 , \; |-1 \rangle_e  \} > \; ,
\end{equation}
where $|0\rangle$ denotes the vacuum, $|1 \rangle$ denotes the
single-photon state in the original
(linear) polarization mode produced by the
down-conversion, and $|-1\rangle$ denotes the single-photon state
in the orthogonal polarization mode,
which is mixed into ${\cal H}_e$ by the polarization
rotator (with complex coefficients $a$,
$b$) placed along the signal beam $x$ (Fig.\,2). The output state can
be written as
\begin{eqnarray}
|\psi\rangle & = &
[\; |0\rangle_u |0\rangle_d |0\rangle_e \nonumber \\
& + &
V f_1 e^{i \phi} \, |1 \rangle_u |0\rangle_d \, (\; a \, |1 \rangle_e
\, + \, b \, |-1\rangle_e \; ) \nonumber \\
& + & V f_2 \, |0\rangle_u |1\rangle_d |1\rangle_e \; ]\, / \, N \; \; ,
\end{eqnarray}
where $|a|^2 + |b|^2 = 1$, and $N$ is the normalization factor
\begin{equation}
N \equiv \sqrt{1 + |V|^2 (|f_1 |^2 + |f_2 |^2 )} \; .
\end{equation}
Notice that the contributions from the signal beam $x$ and from the
signal beam $e$ are coherently superposed in the output state
$|\psi\rangle$ along the ${\cal H}_e$-direction in ${\cal H}$;
this is the key consequence of aligning the two
signal beams.

The experiment consists of monitoring the entangled output
$|\psi\rangle$ at the two single-photon detectors $D_A$ and $D_B$.
For the purposes of our essentially conceptual discussion in this
letter, experimental inaccuracies and noise
(detector inefficiencies, dark-count rates, $\ldots$)
are not relevant, and we will defer their discussion to
a forthcoming paper~\cite{moretocome}. Thus, measurement by the perfect
detector $D_A$ is equivalent to the projection $P_A = P_{\alpha}
\otimes \mathbb{I}_e$, where $\alpha \in {\cal H}_u \otimes {\cal H}_d$
is the vector
$\alpha = (|0\rangle_u |1 \rangle_d
+ i |1\rangle_u |0 \rangle_d )/{\sqrt{2}}$,
and a measurement (click) at $D_B$ is equivalent to the projection
$P_B =  P_{\beta} \otimes \mathbb{I}_e$,
where $\beta = (|1\rangle_u |0 \rangle_d
+ i |0\rangle_u |1 \rangle_d )/{\sqrt{2}}$.
Calculation [using $p_{A,\; B}  =  {\rm Tr}
( P_{A,\; B}  |\psi\rangle\langle \psi | )
 = \| P_{A, \; B} |\psi \rangle \|^2$] shows that
the probabilities $p_A$ and $p_B$ of clicks at detectors $D_A$ and
$D_B$, respectively, are given by
\begin{eqnarray}
p_A & = & \frac{|V|^2}{2 \, N^2} \left[ \,
|f_1 |^2 + |f_2 |^2 \, + \, 2 \, \Re \, ( \,
i \overline{f_1 }f_2 \bar{a} e^{-i \phi }
\, ) \, \right] \; , \nonumber \\
p_B & = & \frac{|V|^2}{2 \, N^2} \left[ \,
|f_1 |^2 + |f_2 |^2 \, - \, 2 \, \Re \, ( \,
i \overline{f_1 }f_2 \bar{a} e^{-i \phi }
\, ) \, \right]  \; .
\end{eqnarray}
The important feature in Eqs.\,(6) is the interference term in brackets
following the real-part sign $\Re$. Notice that the interference is
oscillatory in the controlled phase delay $\phi$ and
depends sensitively on the polarization angles $(a,b)$.

But how does the interference depend on the evolution of
the probe beam $e$ which escapes to infinity? Let
$\rho_{ud} \equiv {\rm Tr}_e |\psi\rangle \langle \psi |$
be the output state projected on the ``laboratory"
Hilbert space ${\cal H}_u \otimes {\cal H}_d$.
It is straightforward to show that
the detection probabilities $p_{A,\; B}$ can be
alternatively computed via the expressions $p_{A, \; B}
= {\rm Tr}_{ud} ( P_{\alpha , \; \beta} \, \rho_{ud})$.
This result is, of course,
valid much more generally: the expectation value of
any observable
$O = O_{ud} \otimes \mathbb{I}_e$ (i.e.\ one local to the
Hilbert space ${\cal H}_u \otimes {\cal H}_d$) depends only on the
reduced state projected on ${\cal H}_u \otimes {\cal H}_d\;$:
\begin{equation}
{\rm Tr}\, [ \, ( O_{ud} \otimes \mathbb{I}_e )\,
|\psi\rangle\langle\psi |\, ] = {\rm Tr}_{ud}\,
[\, O_{ud}\, {\rm Tr}_e |\psi\rangle\langle\psi | \, ] \; .
\end{equation}
Now suppose that the output state $|\psi\rangle$ undergoes a local
quantum evolution (local in the sense that
${\cal E} = {\cal E}_{ud} \otimes {\cal E}_e$)
\begin{equation}
{\cal E} = {\cal E}_{ud} \otimes {\cal E}_e \; : \;
|\psi \rangle
\langle \psi | \mapsto
({\pmb 1}_{ud} \otimes {\cal E}_e )(|\psi \rangle
\langle \psi |) \; ,
\end{equation}
where ${\cal E}_e$ is an arbitrary, completely-positive,
linear quantum map on ${\cal H}_e$-states which
is probability preserving, with Kraus representation:
\begin{equation}
{\cal E}_e  : \rho \mapsto
\sum_{j} E_j \rho {E_j}^\dagger \; , \;\;\;\;\;
\sum_j {E_j}^\dagger E_j = \mathbb{I}_e \; ,
\end{equation}
where $E_j$ are otherwise arbitrary linear operators on ${\cal H}_e$.
For any state $\rho$ on ${\cal H}$ [including the output state
$\rho = |\psi\rangle\langle\psi | $ of Eq.\,(4)], by expanding $\rho$
in the form $\rho = \sum_\mu c_\mu \,
{\rho_{ud}}^{(\mu )} \otimes {\sigma_e}^{(\mu )}$,
$c_\mu \in \mathbb{R}$, it is straightforward
to prove the identity
\begin{equation}
{\rm Tr}_e \, [ \, ({\pmb 1}_{ud} \otimes {\cal E}_e ) \rho \, ]
= {\rm Tr}_e \rho
\end{equation}
for any {\em linear} map ${\cal E}_e$ of the form Eq.\,(9).
In view of Eq.\,(7), Eq.\,(10) is the
expression of causality (no-signaling;
compare Eq.\,(15) below): As long as the evolution of the
probe beam $e$ remains linear and probability-conserving, the
interference pattern of the laboratory beams does not depend on what
happens to $e$. The detection probabilities $p_A$ and $p_B$ are given by
Eqs.\,(6) whether $e$ evolves unitarily, is absorbed in a beam block, or
otherwise gets entangled with the rest of the universe.

By contrast, suppose
that the beam $e$ undergoes a {\em nonlinear},
probability-conserving evolution. As an example, consider
the evolution proposed
in~\cite{horowitzetal} for evaporating black holes, whose
action on any state $\rho \in {\cal H}$ is given by
(see~\cite{bhfs} for a detailed discussion of this map class)
\begin{equation}
{\cal E} \; : \; \rho \longmapsto
\frac{{\pmb 1}_{ud} \otimes {\cal T}_e \; ( \rho)}
{{\rm Tr} [{\pmb 1}_{ud} \otimes {\cal T}_e \; ( \rho)]} \; ,
\end{equation}
where ${\cal T}_e$ denotes the linear transformation (not a quantum map)
${\cal T}_e : \rho_{e} \mapsto T_e  \rho_{e}  {T_e}^{\dagger}$
on states $\rho_e$ of ${\cal H}_e$, and
$T_e : {\cal H}_e \rightarrow {\cal H}_e$
is an arbitrary nonsingular linear transformation.
For simplicity, let us choose $T_e$ in the form
\begin{equation}
T_e = 
  \begin{pmatrix}
   1 & 0 & 0 \cr
   0 & 1 & 0 \cr
   0 & -1 & 1
  \end{pmatrix}
\end{equation}
in the $\{|0\rangle_e , \; |1\rangle_e , \; |-1\rangle_e \}$ basis
of ${\cal H}_e$.
After the incoming state
$\rho=|\psi\rangle \langle \psi|$ is transformed
into ${\rho}^{\ast} \equiv {\cal E}(\rho )$
according to the nonlinear evolution ${\cal E}$
given by Eqs.\,(11)--(12), the probabilities of
detection at the local detectors can be re-calculated using the
equations ${p^\ast}_{A, \; B}
= {\rm Tr} ( P_{A , \; B} \, {\rho^{\ast}})$.
The result for the interference signal $p_A - p_B$ is:
\begin{equation}
{p^\ast}_A - {p^\ast}_B = \frac{2 \, |V|^2}{{N^\ast}^2}\;
\Re \left[ \,
i \overline{f_1 }f_2 \, (\bar{a} - \bar{b}) \, e^{-i \phi }
\, \right] \; ,
\end{equation}
where ${N^\ast}^2 = 1 + |V|^2 [\, |f_2 |^2 + |f_1 |^2 (|a-b|^2+|b|^2
)\, ]$ is the new normalization factor.
Observing a signal like Eq.\,(13) would represent a clean
detection of the nonlinear map $\cal E$ [Eqs.\,(11)--(12)]
by our interferometer, since, e.g., the new null and maximum
of the interference wih respect to the polarization-rotator
angle $\theta \equiv \arctan (b/a)$ are both shifted by $45^{\circ}$
compared to Eqs.\,(6).
In general, the interference signal $p_A - p_B$ as a function of $\phi$ and
$\theta$, the fundamental observable in our proposed experiment,
constitutes a rich 2--D
data set sensitive to almost any nonlinear
evolution map affecting the probe beam $e$.

If black-hole evaporation compels us to treat linearity
as a property to be tested by experiment rather than as an axiom of
quantum mechanics, what
properties must hold for the most general class of
quantum maps governing quantum evolution everywhere?
We now turn briefly to the mathematical description
of this generalized class
of maps. Full details will be found
in the forthcoming~\cite{moretocome}.

Given a bi-partite quantum system $AB$ with
(finite-dimensional) Hilbert space ${\cal H}
\equiv {\cal H}_A \otimes {\cal H}_B$, let $W({\cal H})$
be the real vector space of symmetric operators on $\cal H$,
and $S({\cal H}) \subset W({\cal H})$ 
the set of all states (positive, symmetric operators of unit trace).
We propose that the set of quantum maps $\M ({\cal H}) $ consists of
all smooth maps
${\cal E}_{AB} : W({\cal H}) \rightarrow W({\cal H})$
which map $S({\cal H})$ into $S({\cal H})$ (conserve probability)
and satisfy the {\em locality condition}
\begin{equation}
{\cal E}_{AB} (\rho \otimes \sigma ) = {\cal E}_A
(\rho ) \otimes {\cal E}_B (\sigma ) \; 
\end{equation}
for {\em all} $\rho \in S({\cal H}_A )$
and $\sigma \in S({\cal H}_B )$, where ${\cal E}_A$
and ${\cal E}_B$ are {\em fixed} quantum maps in $\M ({\cal H}_A )$
and $\M ({\cal H}_B )$ (called the {\em local components}
of ${\cal E}_{AB}$) that depend only on ${\cal E}_{AB}$~\cite{ftnote4}.
The condition for a map
${\cal E}_{AB} \in \M$ to be signaling (non-causal) is precisely
that for {\em some} $\rho_{AB} \in S({\cal H})$
\begin{equation}
{\rm Tr}_B \, [ \, {\cal E}_{AB} (\rho_{AB}) \, ]
\neq {\cal E}_A \, [ \, {\rm Tr}_B ( \rho_{AB}  ) \, ] \; ,
\end{equation}
where ${\cal E}_A$ is the local $A$-component of ${\cal E}_{AB}$.
It is easy to prove using Eq.\,(14) that all local linear ${\cal E}_{AB}$
are causal [non-signaling; cf.\ Eq.\,(10)]. Note also
that locality [Eq.\,(14)] explains why phase-coherent
entanglement of
$AB$ is essential to detect any non-causal influence of ${\cal E}_{AB}$
at $A$ when, for example, $B$ is inside the
event horizon and $A$ is in the
exterior region of a black hole:
Any system (e.g., starlight) entangled with the external world
will give rise to a decohered input state having the product form $\rho_{AB}
= \rho_A \otimes \sigma_B$, and evolution of such product states
cannot satisfy the signaling
condition Eq.\,(15) because of the locality constraint Eq.\,(14).
Experiments must carefully preserve
phase-coherence of entanglement (as proposed in Fig.\,2)
to be able to detect signaling.

If we denote the class of (local, completely positive)
linear maps by $\fL \subset \M
({\cal H})$, the complement (nonlinear maps) by $\NL= \M
\setminus \fL$, and the class of
signaling maps by $\fS$, we have just shown that $\fS \cap
\fL = \{ \}$.
In~\cite{moretocome}, we will give a complete
algebraic characterization of nonlinear maps satisfying
the locality condition Eq.\,(14), and
show that it is straightforward to produce both signaling
and non-signaling examples for maps in $\NL$~\cite{gisinetal};
that is, $\fS$ is a non-empty proper subset of $\NL$.
The evolution map $\cal E$ defined by Eqs.\,(11)--(12) is one example of
a class of local nonlinear maps---proposed by Horowitz and Maldacena
in~\cite{horowitzetal}
to describe quantum evolution through evaporating black holes---that
are signaling (i.e., belong to $\fS$). A detailed
analysis of this class of maps, along with a discussion of the motivation
for them, can be found in~\cite{bhfs}.

Let $A\equiv \{u,d\}$ and $B\equiv \{ e \}$
for the output state Eq.\,(4) as the probe beam $e$
is directed into the event horizon of a
black hole. Suppose the evaporation of the hole leads to
a quantum map ${\cal E}_{AB}$ with
a signaling nonlinear component.
Would the nonlinearity cause a detectable
shift in the interference patterns of $u$ and $d$?
Quantum field theory teaches us
that the evolution of $|\psi\rangle$ is
described by ${\cal E}_{AB}$ when and only when
the subsystems $A$ and $B$ are contained in a partial
Cauchy surface $\Sigma_{AB}$
(i.e.\ a spacelike surface no causal curve intersects
more than once). The blue diagram in Fig.\,1 depicts such a surface
$\Sigma_{AB}$ for the causal geometry of the proposed experiment.
If the ultimate causal structure of the evaporating
quantum black hole remains the same
as given by the classical metric (Fig.\,1),
the singularity is a final boundary, $e$ will
propagate unitarily before it disappears into the singularity,
and no signal will be produced (effectively
unitary ${\cal E}_{AB}$).
If, on the other hand,
$e$ re-emerges as Hawking radiation following evaporation
[i.e.\ if the singularity is effectively a part of $H^{+}(\Sigma
)$ in the quantum spacetime], then a detectable signal will result.
Conversely, the likely null outcome of the experiment
can be used to place precise upper limits on
the strength of any signaling nonlinear component in the
effective quantum map ${\cal E}_{AB}$ of evaporating black
holes. An easier to obtain, but perhaps less interesting,
result of the
experiment would be to place novel limits on possible
nonlinearities~\cite{nlrefs,weinberg} in the quantum
evolution of the probe beam $e$ as it propagates through free space.

Environmental decoherence of the probe beam
at large distances (as well as possible
rapid fluctuations of the putative nonlinearities inside black-holes)
places fundamental limits
on the visibility of the interference signal in our proposed experiment;
a detailed analysis of these limits will be given in~\cite{moretocome}.

The research described in this paper was carried out
at the Jet Propulsion Laboratory under a contract with the National
Aeronautics and Space Administration (NASA), and
was supported by grants from NASA and the Defense
Advanced Research Projects Agency.


\begin{thebibliography}{5}
\bibitem{ftnote2} For maps sending pure states to pure
 states, linearity (plus probability conservation) is equivalent
 to unitarity.
\end{thebibliography}
\end{document}